\documentclass{article}

\usepackage{graphicx,amsmath,amssymb,chemarr,subfigure}

\newcommand{\beq}{\begin{equation}}
\newcommand{\eeq}{\end{equation}}
\newcommand{\beqn}{\begin{eqnarray}}
\newcommand{\eeqn}{\end{eqnarray}}

\begin{document}

\title{The macroscopic effects of microscopic heterogeneity}
\author{Andrew Mugler$^1$ and
	Pieter Rein ten Wolde$^{1,}$\thanks{Correspondence: tenwolde@amolf.nl}
\vspace{.15in} \\
\small $^1$FOM Institute AMOLF, Science Park 104, 1098 XG Amsterdam, The Netherlands}
\date{}
\maketitle

\begin{abstract}
Over the past decade, advances in super-resolution microscopy and particle-based modeling have driven an intense interest in investigating spatial heterogeneity at the level of single molecules in cells.  Remarkably, it is becoming clear that spatiotemporal correlations between just a few molecules can have profound effects on the signaling behavior of the entire cell.  While such correlations are often explicitly imposed by molecular structures such as rafts, clusters, or scaffolds, they also arise intrinsically, due strictly to the small numbers of molecules involved, the finite speed of diffusion, and the effects of macromolecular crowding.  In this chapter we review examples of both explicitly imposed and intrinsic correlations, focusing on the mechanisms by which microscopic heterogeneity is amplified to macroscopic effect.
\end{abstract}

\begin{figure}
\centering
\includegraphics[width=3.5in]{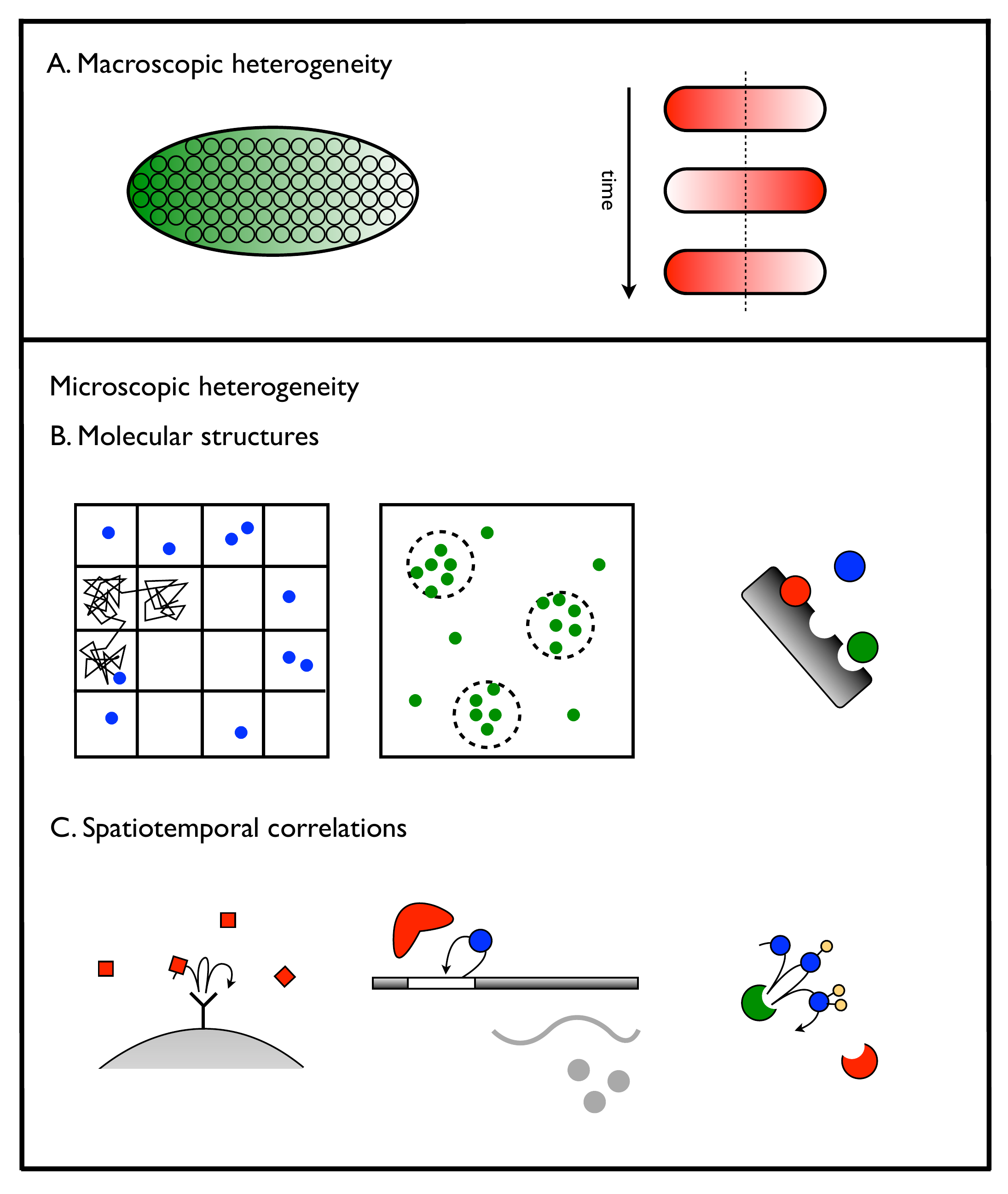}
\caption{Spatial heterogeneity is prominent at both macroscopic and microscopic scales in cellular systems.  (A) Macroscopic heterogeneity: across many cells in an embryo, protein gradients guide segmentation (left); while within single bacterial cells, an oscillating protein gradient guides division (right).  (B) Microscopic heterogeneity: molecular structures, including membrane partitions (left), membrane microdomains (middle), and scaffold proteins (right), introduce microscopic heterogeneities that can have cell-wide effects.  (C) Spatiotemporal correlations resulting from rapid rebinding events can also have pronounced effects on cellular responses.  Here, ligand molecules rapidly rebind to a receptor (left), a transcription factor rapidly rebinds to a DNA operator site, outcompeting an RNA polymerase (middle); and a substrate molecule rapidly rebinds to a kinase to become doubly phosphorylated (right).}
\label{fig:processes}
\end{figure}

It has long been understood that cells perform key functions by exploiting or actively maintaining heterogeneous spatial structures.  Many of these heterogeneities exist at the length scale of the cell or larger.  For example, embryonic segmentation is guided by the formation of protein gradients that stretch from one end of the embryo to the other \cite{Wartlick2009}.  Bacterial cell division is aided by the periodic localization of antagonistic proteins to either cell pole \cite{Lutkenhaus2002}.  These examples are characterized by the intuitive feature that a space-dependent function is performed using a spatially heterogeneous process (Fig.\ \ref{fig:processes}A).

Recent years have seen the discovery of more subtle processes that are homogeneous on cellular scales but nonetheless heterogeneous on molecular scales.  These processes share the remarkable property that microscopic heterogeneities at the level of several molecules can have macroscopic functional consequences at the level of the entire cell.  These heterogeneities are often rooted in molecular-scale structures, such as cytoskeletal partitions \cite{Kusumi1996, Machta2011} or lipid domains \cite{Pike2009, Lingwood2010} on the plasma membrane and molecular clusters \cite{Plowman2005, Sourjik2010}, scaffold proteins \cite{Dard2006} or other molecular complexes (Fig.\ \ref{fig:processes}B).  The ability for such small spatial structures to induce cell-wide changes relies on an intricate network of signaling interactions, such that molecular spatiotemporal correlations can be amplified to produce a macroscopic response.  In this respect, microscopic heterogeneities are functionally important not for their spatial structure per se, but as an additional degree of freedom in signaling computations \cite{Kinkhabwala2010}.

While molecular structures explicitly impose microscopic heterogeneity, recent studies have revealed that microscopic heterogeneity also arises intrinsically, due to spatiotemporal correlations between individual molecules.  Intrinsic heterogeneity is strictly a consequence of small numbers of molecules, the finite speed of diffusion, and the crowded intracellular environment, but when coupled to a nontrivial signaling network, it can have dramatic effects at the cellular level. The predominant mechanism by which spatiotemporal correlations are introduced is via a rebinding process, whereby one molecule rapidly rebinds to another before diffusion can take it far enough away to effectively rejoin the pool of other molecules (Fig.\ \ref{fig:processes}C).  In some cases, the effect of these rebinding events can be captured by renormalizing the parameters governing a system that is otherwise treated as well mixed.  In other cases, however, these rebinding events can place the system in a new dynamic regime or enable a new signaling pathway, leading to a cellular response that is qualitatively different than that of a well-mixed system.  The effects of microscopic heterogeneity are therefore critically related to the signaling network that underlies the molecular interactions.

From a theoretical perspective, it has become clear that the different scales of spatial heterogeneity require different types of models.  Heterogeneity at the cellular scale and larger is well captured by reaction-diffusion models, which treat molecular concentrations as continuous variables that vary in space and time \cite{Wartlick2009, Howard2005}.  Heterogeneity at the molecular scale, on the other hand, demands probabilistic models, which respect the fact that molecules exist in integer numbers.  Spatial variation is then typically treated in one of two ways.  Several methods have been developed for simulating the reaction-diffusion master equation \cite{Hattne2005, Lis2009}, which divides the system into compartments in which molecules are taken to be well mixed.  Alternatively, so-called particle-based modeling techniques, such as lattice modeling \cite{Mugler2012}, Brownian dynamics \cite{Andrews2004, Plimpton2005}, or the exact Green's Function Reaction Dynamics (GFRD) \cite{Takahashi2010}, track the position and time of every molecule in the system.  Importantly, spatiotemporal correlations between individual molecules are often only captured by particle-based models, even though they can produce dramatic effects at cellular scales, where one might naively deem a reaction-diffusion model appropriate \cite{Mugler2012, Takahashi2010}.

In this chapter, we focus on the macroscopic effects of microscopic heterogeneity, reviewing recently studied systems in which heterogeneities at the molecular level give rise to nontrivial effects at the cellular level. We first discuss several systems that exploit molecular structures, such as microdomains, clusters, and scaffolds, to enhance or otherwise change an input-output response.  We then describe several examples of systems that are macroscopically homogeneous but whose response is nonetheless affected both quantitatively and qualitatively by molecular rebinding events.  Throughout, we highlight the importance of both particle-based modeling and experimental techniques with molecular resolution.  The examples presented in this chapter reflect the past decade's increased interest in spatial heterogeneity at the molecular scale.

\section*{Molecular structures modulate cellular responses}

The most straightforward way in which cells maintain molecular heterogeneity is via long-lived molecular structures, such as cytoskeletal compartments, molecular clusters or macromolecules, and scaffolding proteins.  Many structures that play an active role in signaling are localized to the plasma membrane of the cell, as the membrane provides the entry point for detection of environmental stimuli.  In fact, the membrane itself is highly dynamic and complex \cite{Grecco2011}, such that membrane heterogeneities modulate cellular signals before they are even relayed to the inside of the cell for further processing.

\subsection*{Membrane heterogeneity: signal modulation at its entry point}

An important example of membrane heterogeneity involves the interaction of the membrane with its underlying actin skeleton.  The skeleton provides a network of ``fences'' that is anchored by transmembrane protein ``pickets,'' producing a grid of compartments in which membrane-bound proteins are transiently trapped \cite{Kusumi1996} (Fig.\ \ref{fig:processes}B, left).  The presence of these compartments is inferred from single-molecule tracking experiments, which show periods of simple Brownian diffusion interrupted by sudden large hops \cite{Kusumi2005, Kusumi2010}.  At $50$$-$$300$ nm wide, these compartments are only ten or a hundred times larger than the molecules they trap, and because the time to diffuse through them ($\sim$$150$ $\mu$s) is two orders of magnitude faster than the residence time within them ($\sim$$15$ ms), they act as temporary reaction chambers \cite{Kusumi2005}.  Such transient trapping has nontrivial effects on the properties of the propagated signal.  For example, trapping results in less frequent collisions between molecules in different compartments, but more frequent collisions between molecules in the same compartment, such that diffusion-limited reactions occur in rare but potent bursts \cite{Kalay2012}.  In a similar way, trapping enhances oligomerization, which is then further enhanced by the fact that the probability to hop to a neighboring compartment decreases with molecule size \cite{Grecco2011}.  Thus, signals that are initiated by association reactions or oligomer formation can be highly modulated by the presence of compartments on the membrane.

The composition of the membrane is also richly heterogeneous because the lipids that comprise the membrane are not completely mixed.  Instead, regions enriched in glycosphingolipids and membrane proteins, often called lipid rafts or microdomains, transiently assemble and float within the surrounding lipid bilayer \cite{Pike2009, Lingwood2010}.  Indeed, super-resolution microscopy has uncovered complexes of glycosphingolipid and cholesterol existing on $\sim$$10$$-$$20$ ms timescales and $<$$20$ nm length scales \cite{Eggeling2009}.  Rafts are thought to serve as platforms for signal transduction \cite{Simons2000}.  In particular, they have been connected to the observation that certain membrane-bound proteins form clusters in a cholesterol dependent manner \cite{Kholodenko2010}.  For example, immuno-electron microscopy experiments have revealed that Ras, a membrane-bound eukaryotic protein implicated in a variety of phenotypic responses including oncogenesis, forms cholesterol-dependent clusters of roughly seven molecules \cite{Plowman2005, Prior2003} (Fig.\ \ref{fig:processes}B, middle).  Single-molecule tracking experiments confirm that Ras molecules exhibit suppressed diffusion upon activation \cite{Murakoshi2004, Lommerse2005}, suggesting further that clustering correlates with molecules' activation state \cite{Hancock2005}.  Ras plays a critical role in transducing extracellular signals: once activated in response to receptor stimulation, Ras initiates a mitogen-activated protein kinase (MAPK) cascade that ultimately leads to transcription.  Theoretical studies have shown that the organization of Ras in clusters improves the fidelity of this transduction by converting a graded response to a digital one \cite{Tian2007} and thereby reducing the intrinsic noise \cite{Gurry2009}.  The discrete nature of the molecular structure therefore translates to a signaling advantage at the level of the entire cell.

\subsection*{Clusters and scaffolds: competing effects on mean responses}

The discovery of molecular clusters on eukaryotic membranes resonates with the long standing observation that the receptors that initiate chemotaxis form clusters on bacterial membranes \cite{Sourjik2010}.  The prevalence of membrane clustering has thus launched considerable efforts to quantify its effect on signal detection, amplification, and subsequent propagation.  With regard to detection, for example, it has been recognized that receptor clustering promotes the repeated reversible binding of a ligand molecule, a phenomenon termed serial ligation.  Serial ligation was first studied in the context of T cell activation, where it was proposed to account for the high sensitivity of T cells to low numbers of ligand molecules \cite{Valitutti1995, Wofsy2001}.  Brownian dynamics simulations have been used more recently to investigate the role of serial ligation for clustered receptors, revealing that rather than amplifying or attenuating signals on average, repeated ligand binding affects the statistics: localized detection events occur in rapid succession, with long intervals in between \cite{Andrews2005}.  Such spatiotemporal correlations indeed have the potential to boost the sensitivity to low ligand concentrations, at the cost of increasing the noise in the time-averaged detected signal \cite{Andrews2005}.

With regard to amplification, it has been demonstrated both theoretically and experimentally that incoming signals can be amplified by coupling of receptors to each other within a cluster.  Such coupling, in which the ligand-induced activation of one receptor biases the activation of neighboring receptors, has been modeled using techniques from statistical mechanics, originally by Monod, Wyman, and Changeux \cite{Monod1965} and Koshland, N{\'e}methy, and Filmer \cite{Koshland1966}, and more recently by Bray and Duke \cite{Bray2004}.  Indeed, modeling \cite{Duke1999, Mello2003} and experiments in bacteria \cite{Sourjik2004} have shown that receptors operate in a highly cooperative manner, mimicking the behavior of allosteric proteins, ultimately leading to enhanced sensitivity to the incoming ligand signal.

\begin{figure}
\centering
\includegraphics[width=3.5in]{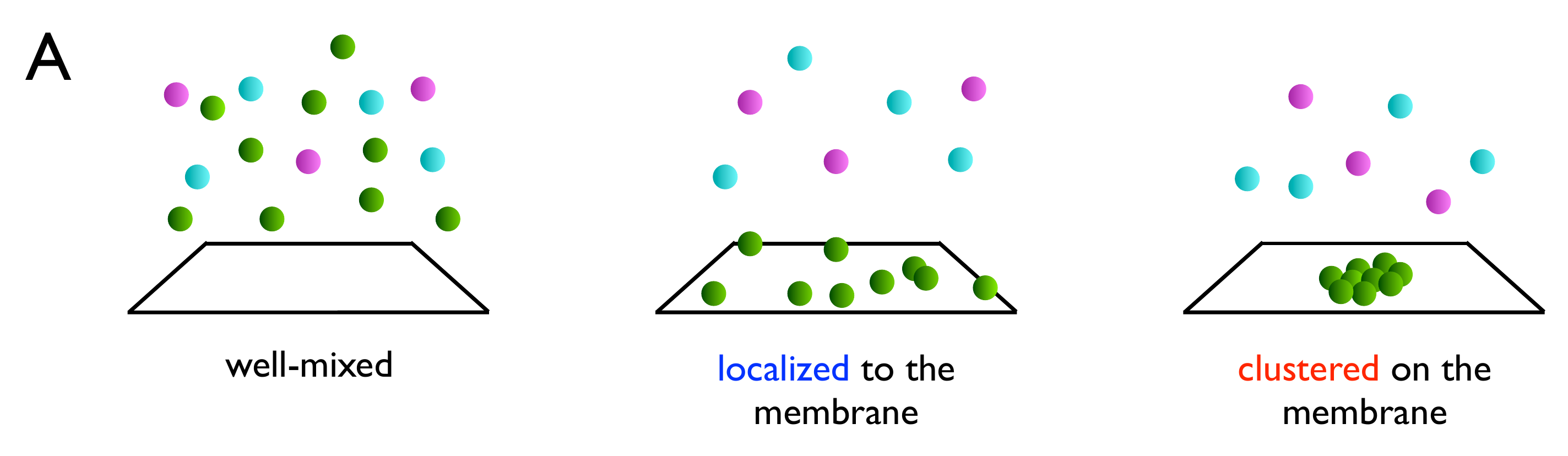}
\includegraphics[width=3.5in]{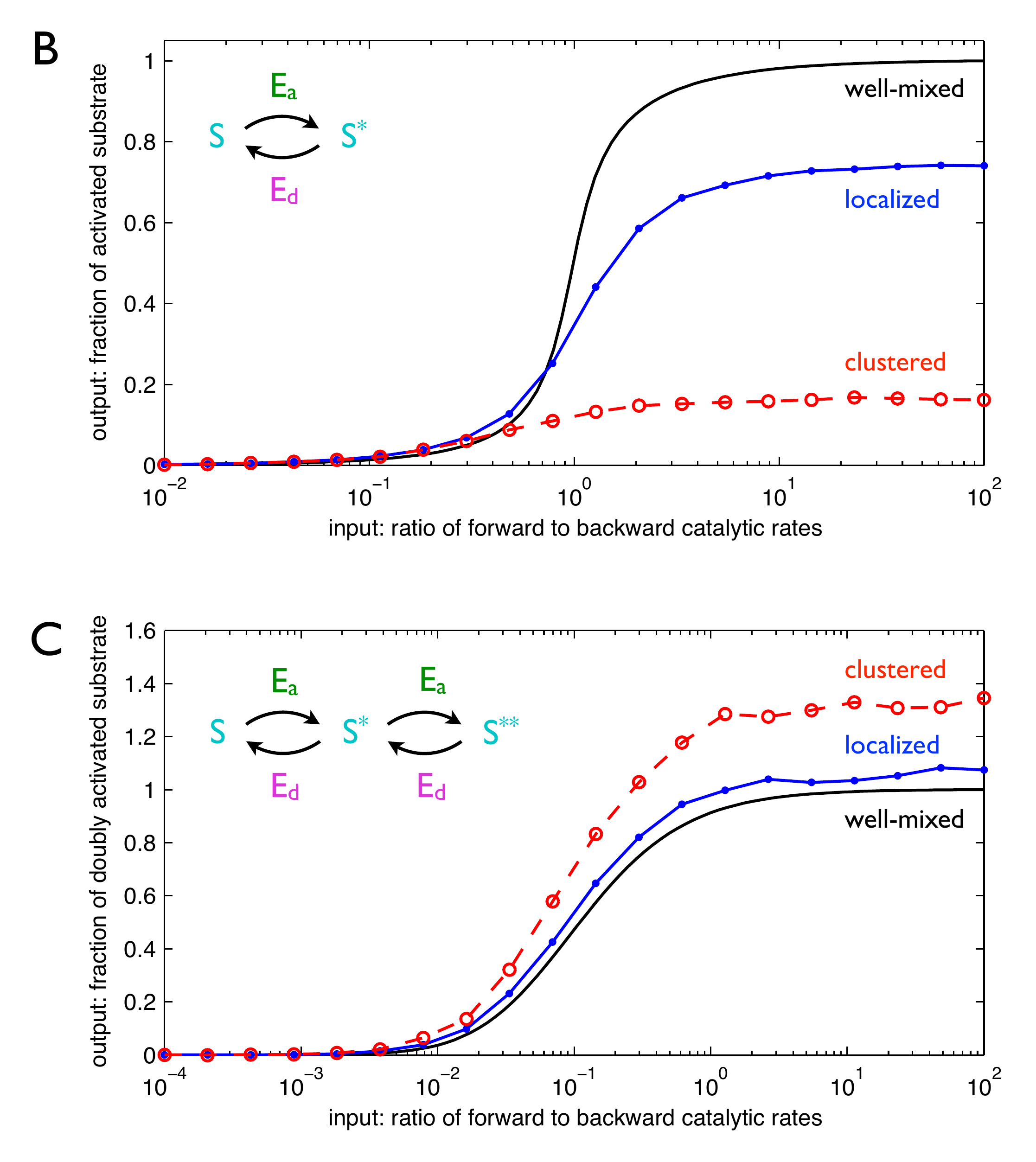}
\caption{Membrane clustering can reduce or enhance a signaling response.  (A) A substrate (cyan) is covalently modified by an activating enzyme (green) and a deactivating enzyme (magenta).  Activating enzyme molecules are either well-mixed in the bulk (left), localized to the membrane randomly (middle), or clustered on the membrane (right).  (B) In a single-modification network, the response is reduced with clustering, since clustering reduces the target size as seen from the bulk.  (C) In a double-modification network, on the other hand, the response can be enhanced with clustering, since clustering promotes rapid rebinding of substrate molecules.  The output is normalized by the maximum value of the well-mixed curve.  Data are from Fig.\ 2 of \cite{Mugler2012}; cluster size is $25$ molecules.}
\label{fig:cluster}
\end{figure}

Finally, protein clustering not only allows amplification of the incoming signal, it also changes the propagation of the signal from the receptors at the membrane to the downstream signaling pathway in the cytoplasm.  In the case of both the Ras system and the bacterial chemotaxis system, the clustered proteins drive a covalent modification network often called the push-pull network, a motif that is also ubiquitous in intracellular signaling \cite{Sauro2004}.  In a push-pull network, a substrate is activated and deactivated by two antagonistic enzymes (Fig.\ \ref{fig:cluster}B, inset).  Lattice simulations have shown that localizing the activating enzyme to clusters on the membrane leads to competing effects which can either reduce or enhance the mean response \cite{Mugler2012}. On the one hand, clustering the activating enzyme molecules presents a smaller target to a substrate molecule in the bulk.  Clustering thus increases the average time for a substrate molecule to find an activating enzyme molecule by diffusive search, an effect that can reduce the response (Fig.\ \ref{fig:cluster}B).  On the other hand, clustering promotes rapid rebinding of a substrate molecule to an activating enzyme or its neighbors.  In a double-modification push-pull network (Fig.\ \ref{fig:cluster}C, inset) in which the substrate molecules must rebind to an enzyme to become doubly activated (a so-called distributive mechanism), rapid rebinding can overcome the effect of the target size reduction and instead enhance the response (Fig.\ \ref{fig:cluster}C).  This example shows explicitly that the effects of molecular heterogeneity are critically dependent on the underlying signaling network.

It is important to emphasize that signal detection by the reversible binding of a ligand molecule to receptors constitutes an equilibrium process.  Changing the spatial distribution of receptors therefore changes the statistics of the response but leaves the mean unaffected.  The reason is that the mean occupancy follows directly from the partition function for an equilibrium process, which in this case is independent of spatial configuration \cite{Mugler2012}.  In contrast, signal propagation is quite often a non-equilibrium process.  Indeed, Fig.\ \ref{fig:cluster} demonstrates that when molecular heterogeneity is coupled to a non-equilibrium network such as the push-pull network, the mean response changes \cite{Mugler2012, vanAlbada2007}.  Downstream signaling from membrane clusters therefore provides a key example of how molecular heterogeneity can modulate mean cellular responses when coupled to a non-equilibrium signaling network.

The observation that molecular structures can have competing effects on mean responses has also been made with regard to scaffolding proteins.  Scaffolds are docking proteins that often serve to assemble the appropriate molecular components of signaling networks (Fig.\ \ref{fig:processes}B, right), including MAPK cascades \cite{Kolch2005, Dard2006}.  Lattice Monte Carlo simulations of a MAPK cascade have shown that scaffolds amplify signals that would otherwise be attenuated or attenuate signals that would otherwise be amplified \cite{Locasale2007}.  Amplification occurs when phosphatase activity is high and therefore the active state of the kinase is short-lived.  In this case, scaffolds promote the rapid phosphorylation of downstream kinases that neighbor each other on the scaffold, thus preserving a signal that in solution would be attenuated by the phosphatases.  Conversely, attenuation occurs when phosphatase activity is low and therefore the active state of the kinase is long-lived.  In this case, scaffolds limit the number of downstream molecules that a kinase can phosphorylate to those that are bound to the scaffold, whereas the kinase could instead phosphorylate many more molecules by free diffusion.  The opposing effects illustrate that the properties conferred to a signaling network by a molecular structure depend nontrivially on the biochemical parameters, a feature that is also true for signals stemming from membrane clusters \cite{Mugler2012}.

\subsection*{Macromolecules: the effect of dimensionality}

While the membrane imposes a roughly two-dimensional geometry, many macromolecules and polymers inside cells, such as DNA, microtubules, and actin filaments, are quasi-linear and thus impose a one-dimensional geometry.  Since the properties of Brownian diffusion are critically dependent on dimension \cite{Redner2001}, many have wondered if the linearity of macromolecules plays a significant role in the signaling computations in which they take part.  The question of dimensionality and signaling is perhaps best exemplified by the problem of a transcription factor protein locating the operator site on a strand of DNA in order to regulate the expression of a gene.  The search can occur by a process known as facilitated diffusion, in which three-dimensional (3D) diffusive motion in the cytoplasm is combined with periods of one-dimensional (1D) diffusive sliding along the DNA \cite{Berg1981}.

While it would seem that a combined 1D--3D search strategy might have very different statistical properties than 3D diffusion alone, recent work has revealed that, surprisingly, this is not necessarily the case.  For example, one may consider the mean time for a protein to find the operator site, which is approximately given by $(L/\lambda)(\lambda^2/D_1+r^2/D_3)$ \cite{Hu2006}.  Here $L$ is the total length of the DNA, $\lambda$ is the average distance over which the protein slides along the DNA before it dissociates, $r$ is the characteristic distance in 3D space between two segments on the DNA, $D_1$ is the diffusion coefficient for sliding, and $D_3$ is the diffusion coefficient in the cytoplasm.  This expression has a clear interpretation \cite{Hu2006}: $\lambda^2/D_1$ is the sliding time, $r^2/D_3$ is the time spent on 3D diffusion, the sum of these terms is thus the time to perform one round of sliding and diffusion, and $L/\lambda$ is the total number of rounds needed to find the target.  One immediately sees that the search time is minimized when $\lambda=r\sqrt{D_1/D_3}$, corresponding to equal time spent sliding in 1D and diffusing in 3D.  A key insight, however, is that on length scales beyond the mean sliding length, the motion essentially appears as 3D diffusion \cite{Gerland2002}.  In fact, theoretical estimates \cite{vanZon2006} and recent experiments \cite{Stanford2000, Gowers2005} show that the sliding length is quite short ($\lambda \lesssim 15$ nm), comparable to the typical size of the protein itself.  In the end, then, facilitated diffusion appears very similar to 3D diffusion alone on the length and timescales that are relevant for the behavior of the network.

One might further expect that facilitated diffusion would reduce the noise in gene expression caused by the diffusive arrival of transcription factors.  This expectation stems from the intuition that noise should decrease with the size of the target and that permitting 1D diffusion should dramatically increase the effective target size \cite{Hu2006}.  However, recent theoretical work has shown that that this intuition is incorrect; instead, the noise from facilitated diffusion is roughly equal to the noise from 3D diffusion alone \cite{Tkacik2009}.  The reason is that the increased size of the target is largely canceled by the strong temporal correlations characteristic of 1D diffusion.  Specifically, the variance scales as $1/T$ in 3D but $1/\sqrt{T}$ in 1D, such that integrating over a fixed time $T$ gives fewer independent samples in 1D than in 3D \cite{Tkacik2009}.  Thus, with respect to the noise as well as the mean, the dimensionality reduction introduced by quasi-linear macromolecules can have effects that are far from straightforward.

\section*{Spatiotemporal correlations modulate responses even in homogeneous systems}

When we call a system ``well mixed,'' we often imagine a chemical reaction in a test tube.  In a test tube, the number of molecules involved is huge, allowing concentrations to be treated as continuous variables.  Diffusion (or active stirring) then creates an environment in which the probability of reacting is proportional to the bulk concentrations of reactants.  In cells, however, reactions can be far from well mixed.  The number of molecules of a specific type can be in the tens or hundreds (or even one, in the case of, say, a DNA operator site).  The diffusion of a handful of molecules then can no longer be treated as creating a homogeneous system.  Instead, spatiotemporal correlations between individual molecules give rise to heterogeneity.  This heterogeneity is quite distinct from that explicitly imposed by actively maintained molecular structures.  Indeed, it arises strictly due to the small number of molecules and the finite speed of diffusion.  Nonetheless, just as with molecular structures, correlations caused by this inherent heterogeneity can be amplified by a signaling network, giving rise to appreciable changes over the entire cell.

\subsection*{Rapid rebinding}

Under the assumption that a system is well mixed, the probability that a molecule reacts within any short time window is constant, proportional to the bulk concentration of its reaction partner(s).  A direct consequence of this fact is that the waiting time between reaction events is an exponentially distributed random variable.  In reality, however, molecules must diffuse toward each other in order to react.  The probability of reacting with a particular partner molecule can be heightened simply by the fact that, by virtue of a recent unbinding event or a random diffusive fluctuation, it is nearby.  The result is that a system cannot be well mixed on all timescales.  While on sufficiently long timescales the waiting time will be exponentially distributed, on short timescales it will follow a distribution whose form will include algebraic scalings typical of diffusion-mediated interactions.  This observation immediately identifies the rebinding problem---i.e.\ measuring the waiting times between successive binding events---as a key tool in exploring the continuum between rapid behavior that is spatiotemporally correlated and long-time behavior that is effectively well mixed.

\begin{figure}
\centering
\includegraphics[width=3.5in]{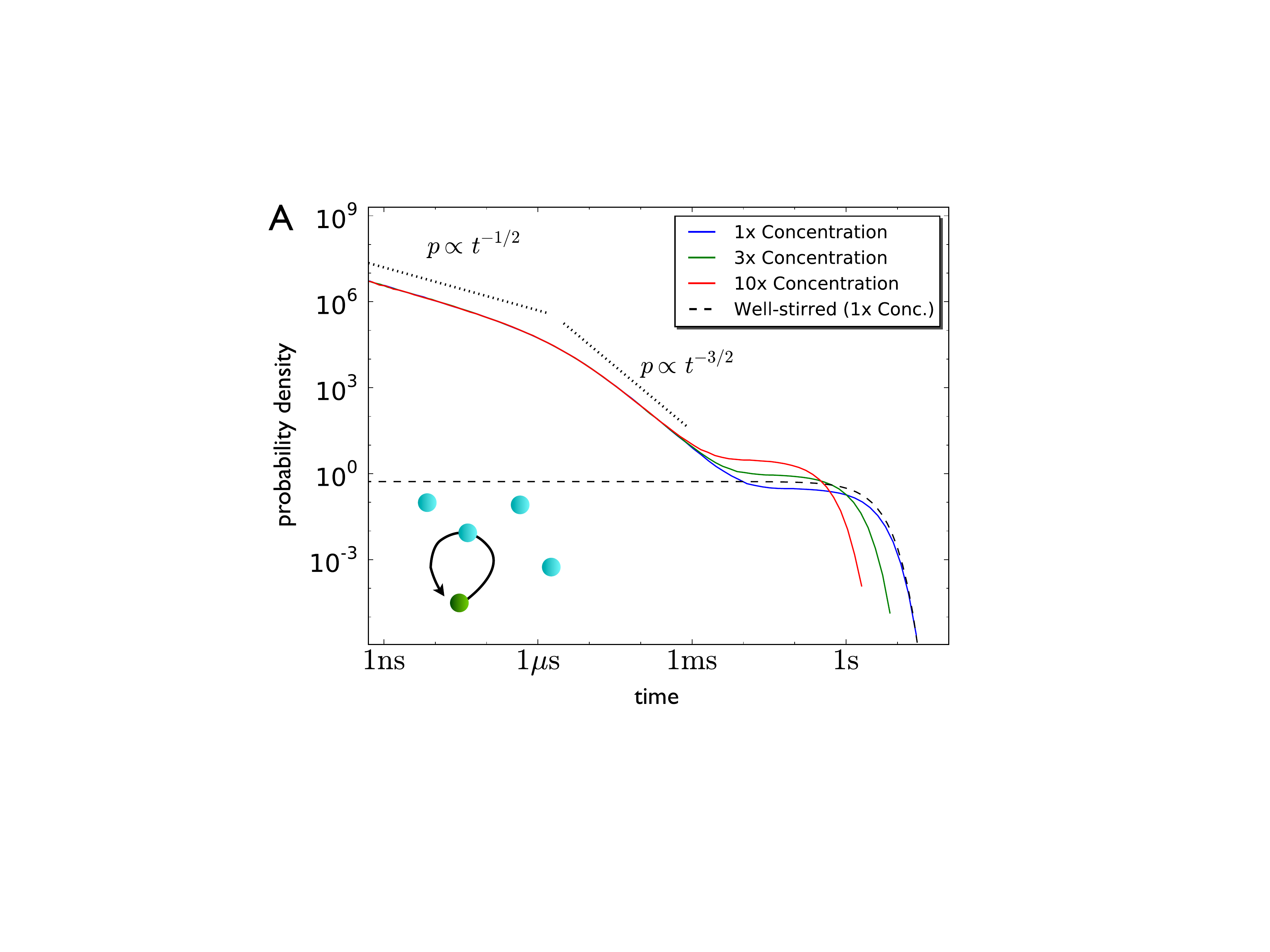}
\includegraphics[width=3.5in]{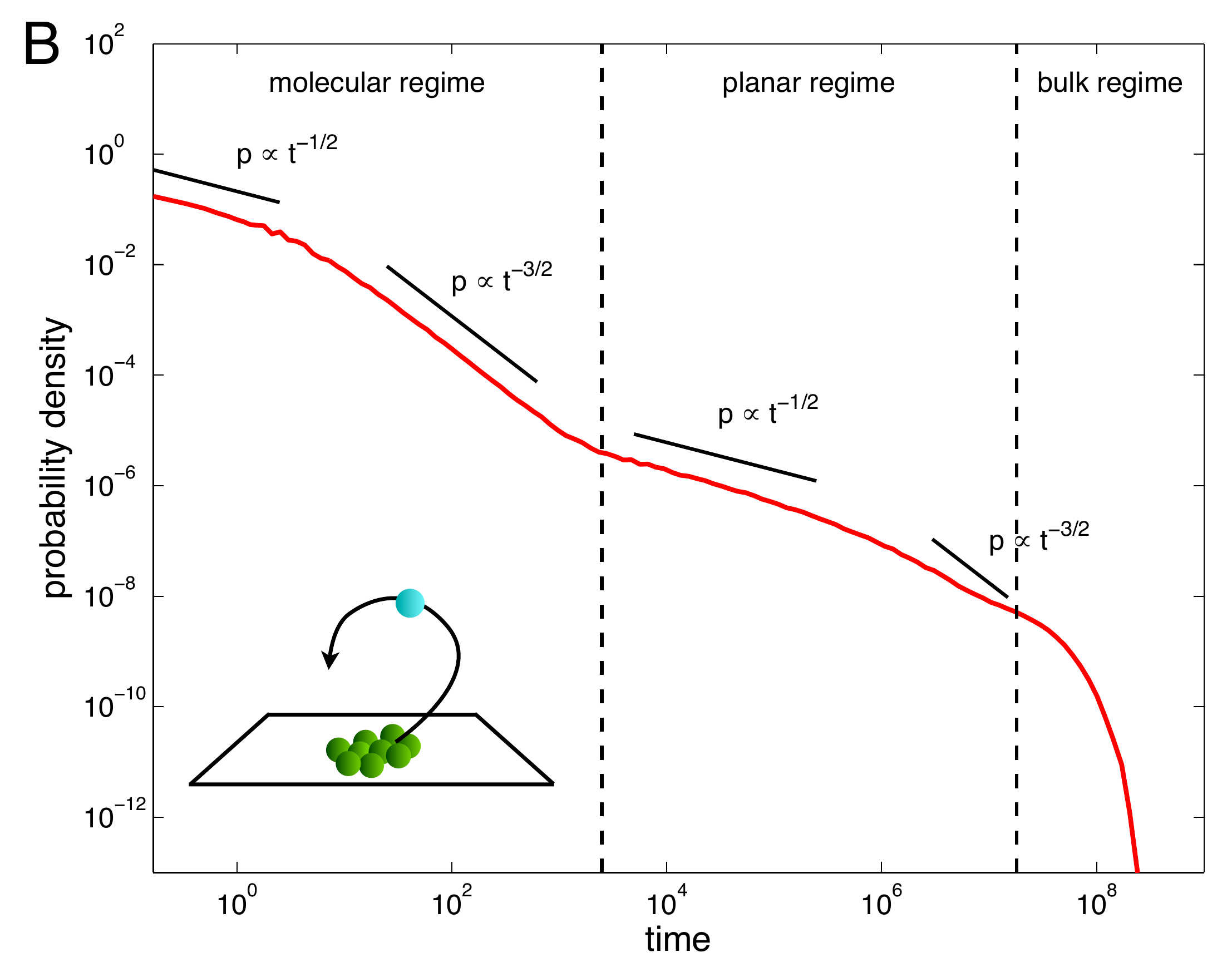}
\caption{The distribution of rebinding times reveals departures from a well-mixed system.  (A) Rebinding in three dimensions.  A single $A$ molecule binds one of $10$ $B$ molecules via $A + B \rightleftharpoons C$.  The concentration $[B]$ is controlled by changing the volume.  The concentration-independent algebraic scalings are due to the rebinding of the same $B$ molecule, whereas the concentration-dependent exponential decay is due to the arrival of a $B$ molecule from the bulk and agrees with the well-mixed prediction.  Data are from Fig.\ 2 of \cite{Takahashi2010}.  (B) Rebinding to clusters on a plane.  A cluster of $100$ $A$ molecules on a plane binds a single $B$ molecule in the three-dimensional bulk via $A + B \rightleftharpoons C$.  The algebraic scalings in the molecular and planar regimes are due to the rebinding of the $B$ molecule to the same cluster and to neighboring clusters (via periodic boundaries), respectively.  The exponential decay in the bulk regime is due to the $B$ molecule reflecting off the wall opposite the plane, effectively randomizing its position and newly arriving from the bulk, as in a well-mixed system.  Parameters are as in Fig.\ 4 of \cite{Mugler2012}, except $\zeta = 6000$ and $\delta=5$; time is scaled by the typical time to diffuse one molecular diameter, $\ell^2/D$.}
\label{fig:rebind}
\end{figure}

The rebinding problem is most simply posed by considering a single $A$ molecule that can react with one of many freely diffusing $B$ molecules in fixed volume according to the reversible reaction $A + B \rightleftharpoons C$.  This problem can be modeled in a spatially resolved way using particle-based modeling techniques; results from a simulation with GFRD \cite{Takahashi2010} are shown in Fig.\ \ref{fig:rebind}A.  Interestingly, both exponential and algebraic regimes are observed in the distribution of rebinding times $t$.  At long times, the distribution is exponential, characteristic of a well-mixed system.  The exponential dependence corresponds to events in which a $B$ molecule unbinds from the $A$ molecule and diffuses far enough away to be effectively randomized within the bulk of other $B$ molecules.  At short times, the distribution is algebraic, following a $t^{-1/2}$ scaling at the shortest times, and a $t^{-3/2}$ scaling at intermediate times.  The algebraic dependence corresponds to events in which a $B$ molecule unbinds from the $A$ molecule but remains in the vicinity and rebinds before escaping to the bulk --- so-called rapid rebinding events.  The shortest times correspond to collision-dominated excursions: these  extremely local (on the order of the molecular diameter) excursions are dominated by many unsuccessful reflections and thus inherit the $t^{-1/2}$ scaling from the Gaussian Green's function of a particle freely diffusing in one dimension \cite{Mugler2012}.  Intermediate times correspond to search-dominated excursions: these longer excursions find the $B$ particle returning to an effectively absorbing boundary, producing the $t^{-3/2}$ scaling characteristic of a three-dimensional random walker returning to an absorbing origin \cite{Mugler2012}.  Importantly, Fig.\ \ref{fig:rebind}A demonstrates that the exponential portion of the rebinding time distribution is a function of the concentration of $B$ molecules, as one would expect for a well-mixed system.  In contrast, the algebraic portion is independent of concentration because it is a local feature arising from the interaction between just two molecules.

The rebinding time distribution is also a useful tool for understanding the spatiotemporal effects of molecular structures, such as the membrane clusters depicted in Fig.\ \ref{fig:cluster}A (right).  Analytic tools \cite{Lagerholm2000, Gopalakrishnan2005, Chevalier2011}, Brownian dynamics simulations \cite{Andrews2005} and lattice simulations \cite{Mugler2012, Gopalakrishnan2005} have been used to study the distribution of times for a particle to rebind to a set of clusters on a plane.  Fig.\ \ref{fig:rebind}B shows that such a distribution also exhibits exponential dependence at long times (the bulk regime) and algebraic scalings at short times.  Here, however, the algebraic regime exhibits two subregimes.  The shortest times (the molecular regime) correspond to rebinding to the same cluster, producing the same $t^{-1/2}$ and $t^{-3/2}$ scalings seen in Fig.\ \ref{fig:rebind}A from collision-dominated and search-dominated excursions, respectively \cite{Mugler2012}.  Intermediate times (the planar regime) correspond to excursions in which the particle diffuses far enough away that the granularity of the clusters disappears, and the plane appears to be uniformly covered with the activating enzyme.  This regime also exhibits $t^{-1/2}$ and $t^{-3/2}$ scalings, corresponding once more to collision-dominated and search-dominated returns, respectively \cite{Mugler2012}.  Importantly, the weight of the distribution within the molecular regime is the key determinant of the strength of rebinding and therefore underlies the enhancement of the mean response for the double-phosphorylation network seen in Fig.\ \ref{fig:cluster}C \cite{Mugler2012}.

\subsection*{Renormalization: integrating out the rebinding}

The fact that rebinding events are so rapid when compared to the timescale of bulk redistribution raises the question of whether rebinding can be captured in a more coarse-grained description \cite{vanZon2006, deRonde2012}.  That is, it is natural to ask whether rebinding can be absorbed into the description of an effectively well-mixed system by a simple rescaling of the binding and unbinding rates.  Indeed, van Zon et al.\ showed \cite{vanZon2006} that in the context of gene regulation, the effect of the diffusion and rebinding of transcription factor proteins to a DNA operator site can be captured in a well-mixed model by renormalization of the rates as follows:
\beqn
\label{eq:kon}
k_{\rm on} &=& \left( \frac{1}{k_a} + \frac{1}{k_D} \right)^{-1}, \\
\label{eq:koff}
k_{\rm off} &=& \left( \frac{1}{k_d} + \frac{K_{\rm eq}}{k_D} \right)^{-1}.
\eeqn
Here,
$k_d$ is the dissociation rate of the protein from the operator site, $k_a$ is the intrinsic association rate (i.e.\ the rate of binding given that the protein is in contact with the operator site), $k_D$ is the diffusion-limited association rate (i.e.\ the rate of a protein finding the operator site from the bulk; in three dimensions $k_D = 4\pi\sigma D$ for binding cross-section $\sigma$), and $K_{\rm eq} = k_a/k_d$ is the equilibrium constant. The rates $k_{\rm on}$ and $k_{\rm off}$ are not the intrinsic association and dissociation rates.  Instead, $k_{\rm on}$ and $k_{\rm off}$ are effective rates, accounting both for the intrinsic activity and for diffusion.  Thus, the result naturally suggests a coarse-grained interpretation of binding as an effectively two-state system, so long as we properly renormalize the binding and unbinding rates as in Eqns.\ \ref{eq:kon} and \ref{eq:koff}.

Interestingly, Eqns.\ \ref{eq:kon} and \ref{eq:koff} have a clear interpretation in terms of the number of rapid rebinding events \cite{deRonde2012}.  The probability of rebinding $n$ times, then exiting to the bulk, is $(1-p)^np$, where $p = k_D/(k_a+k_D)$ is the probability of escaping a reactive origin in three dimensions \cite{vanZon2006}.  The average number of rebinding events is then $N = \sum_{n=0}^\infty n(1-p)^np = (1-p)/p = k_a/k_D$.  Assuming the time to rebind is short compared to the time spent bound $k_d^{-1}$, the total time spent bound can be estimated as $(1+N)k_d^{-1} = k_d^{-1}+K_{\rm eq}/k_D = k_{\rm off}^{-1}$.  We thus arrive at a very useful interpretation of Eqn.\ \ref{eq:koff}: $k_{\rm off}$ is the coarse-grained dissociation rate, renormalized by averaging over the rapid rebinding events.  Because reversible binding is an equilibrium process and therefore obeys the detailed-balance condition $K_{\rm eq} = k_a/k_d = k_{\rm on}/k_{\rm off}$, it follows that the coarse-grained association rate must be renormalized by the same factor $(1+N)$.  Indeed, renormalizing, we reproduce Eqn.\ \ref{eq:kon}: $(1+N)k_a^{-1} = 1/k_a+1/k_D = k_{\rm on}^{-1}$.

\begin{figure}
\centering
\includegraphics[width=6.3in]{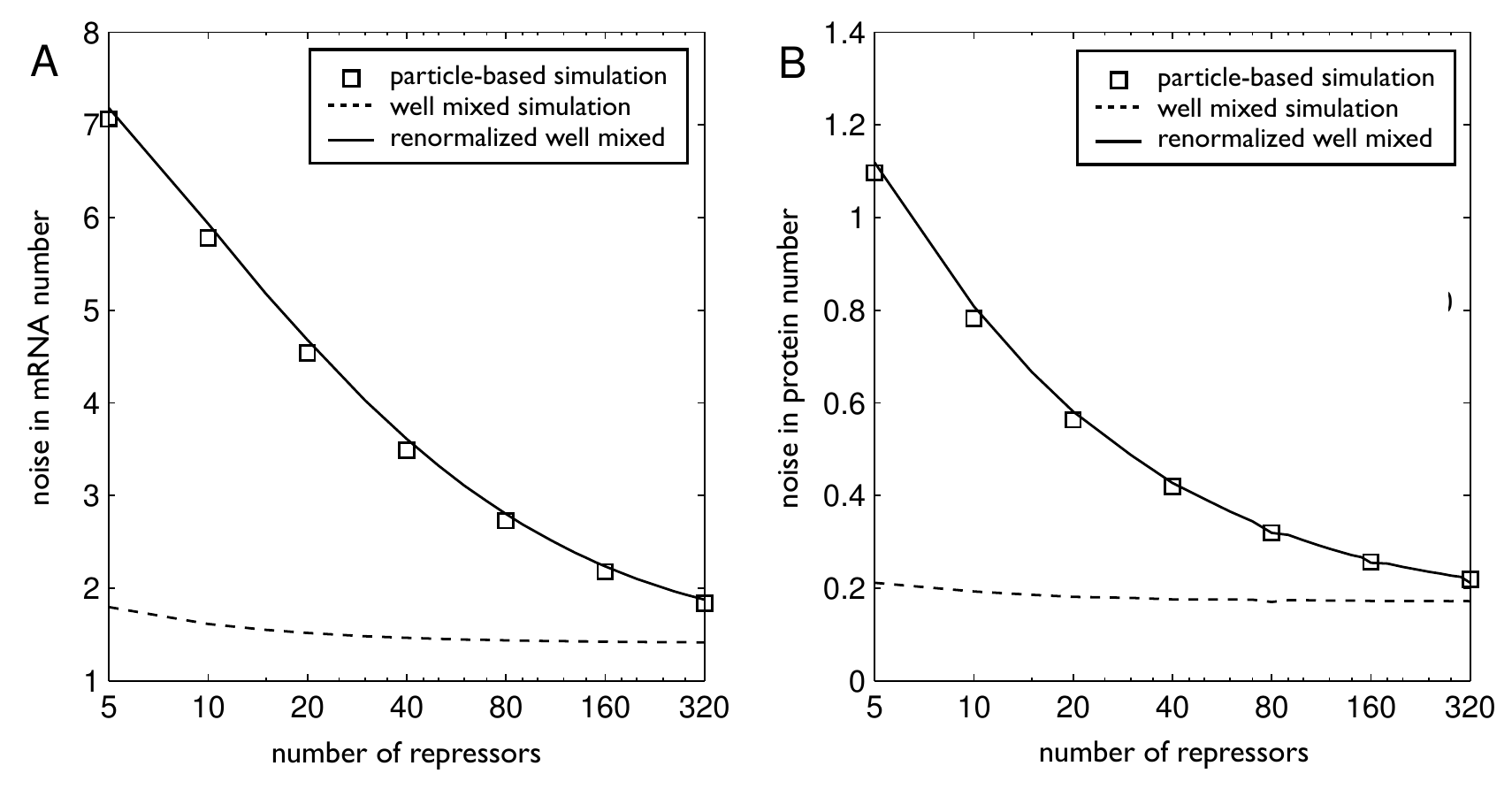}
\includegraphics[width=3.1in]{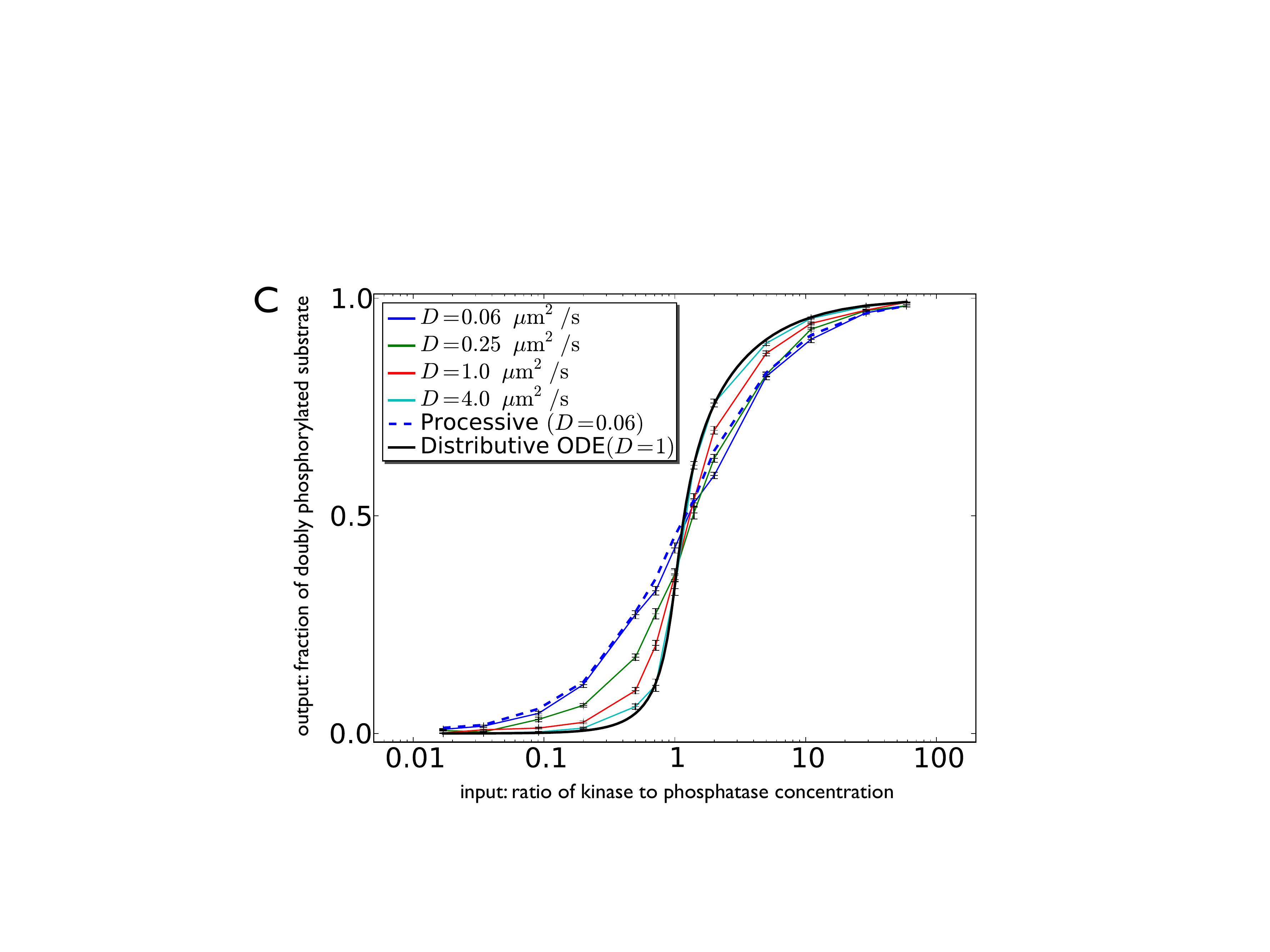}
\includegraphics[width=3.1in]{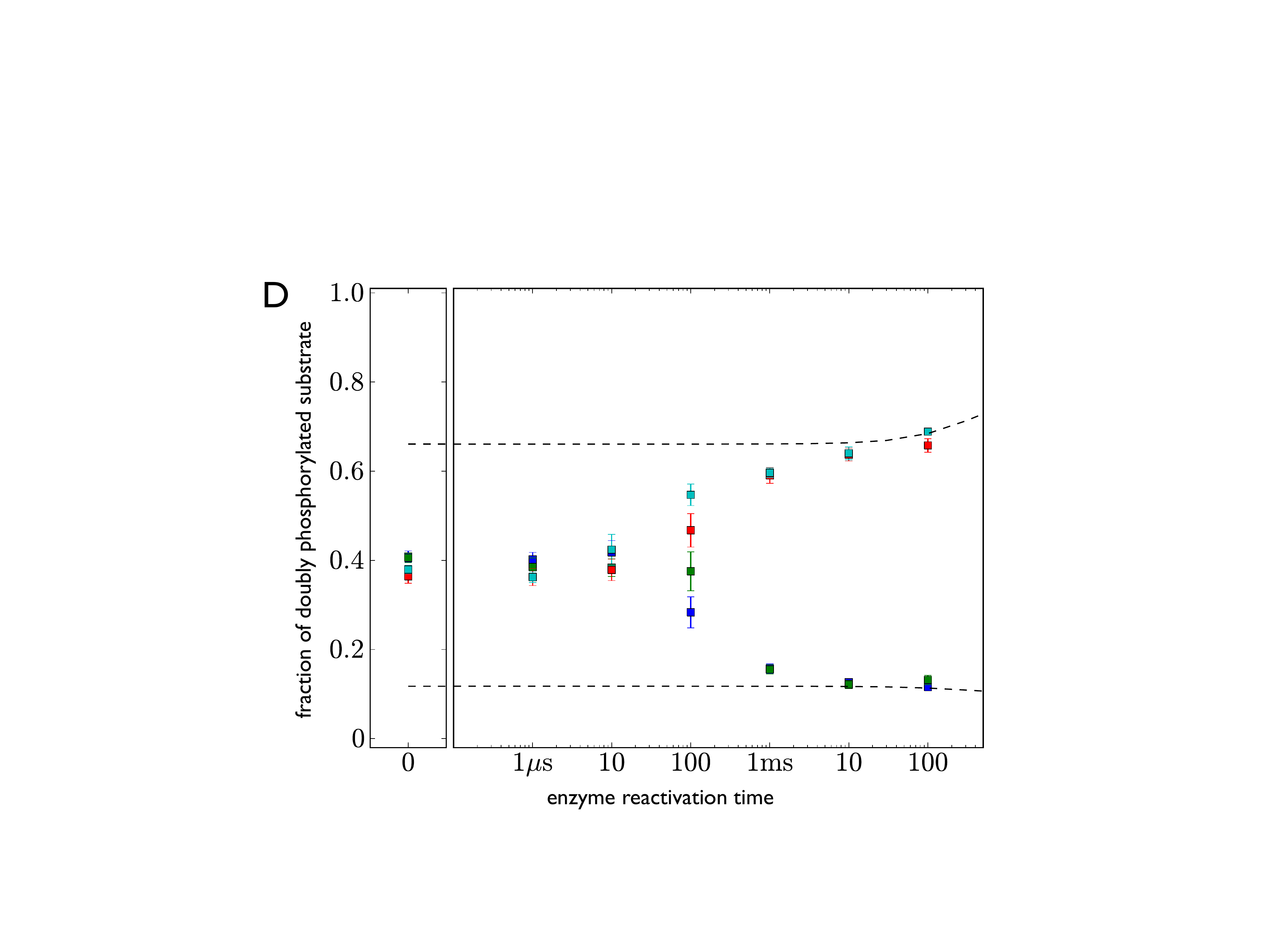}
\caption{Examples in which rebinding can (A, B) and cannot (C, D) be renormalized. (A, B) A transcription factor represses production of mRNA and protein by competing with RNA polymerase for the DNA operator site.  Spatiotemporal correlations between the repressor and the DNA, as captured by the particle-based model, dramatically increase the noise (variance divided by the squared mean) in the mRNA number (A) and protein number (B) beyond that of the well-mixed model.  Because the timescale of repressor rebinding is much faster than that of RNA polymerase binding, the increased noise is well captured by a well-mixed model with rates renormalized by the average number of rebinding events.  Data are from Fig.\ 2 of \cite{vanZon2006}.  (C, D) A distributive dual phosphorylation network, as in one layer of a MAPK cascade.  As the diffusion coefficient decreases, the input-output response becomes less ultrasensitive (i.e.\ less sharp), ultimately approaching that of a processive network (C).  In the bistable regime, as the enzyme reactivation time decreases, the system goes from bistable to monostable (D). In D, the dashed lines show the two bistable states of the well-mixed system, while the blue, green, red, and cyan data correspond to successively larger initial conditions for the fraction of doubly phosphorylated substrate.  It is seen that at enzyme reactivation times greater than $\sim$$100$ $\mu$s, simulations proceed to one of two fixed points, indicative of bistability, while at times less than $\sim$$100$ $\mu$s, they converge to one monostable state; in between, the system is close to a critical point, with concomitant large and long-lived fluctuations.  Loss of ultrasensitivity and bistability constitute qualitative changes in the system behavior that depart from the well-mixed predictions in ways that cannot be captured by renormalization.  Data are from Fig.\ 5 (C) and Fig.\ 6 (D) of \cite{Takahashi2010}.}
\label{fig:renorm}
\end{figure}

The validity of renormalization was demonstrated by showing that this approach not only captures the mean but also the noise in downstream transcription and translation. Particle-based simulations with GFRD \cite{vanZon2006} were used to study the dynamics of a transcription factor that represses transcription by binding to the DNA operator site, thus interfering with the binding of the RNA polymerase (Fig.\ \ref{fig:processes}C, middle).  The effective bound time of the repressor is extended by rapid rebinding events, which in this case can dramatically increase the noise in the production of mRNAs and proteins.  Because these rebinding events occur on a much faster timescale than the binding and unbinding of the RNA polymerase, groups of rebinding events can safely be treated as single binding events in a well-mixed description with the appropriately renormalized rates.  Simulation of transcription and translation using a well-mixed model with these renormalized rates then indeed reproduces with high accuracy the noise in the numbers of mRNAs and proteins (Fig.\ \ref{fig:renorm}A and B).  Renormalization is thus a powerful tool in this context, not only providing a simplified description of the rapid rebinding process but also yielding accurate statistics for its downstream effects.

The treatment of rebinding in a renormalized description has more recently seen favorable comparison with experiments on T cell receptors \cite{Govern2010, Aleksic2010}.  A long-standing puzzle in the field of T cell ligand potency was whether the response of a T cell is governed by how long a ligand molecule stays bound to a receptor, or how often receptors are occupied by ligand molecules on average.  Eqn.\ \ref{eq:koff} illustrates that the effective bound time, once renormalized for rebinding, actually depends on both.  That is, the effective bound time $k_{\rm off}^{-1}$ depends on both the intrinsic bound time $k_d^{-1}$ and the equilibrium constant $K_{\rm eq}$, which dictates the occupancy.  Indeed, using the two-dimensional analog of Eqn.\ \ref{eq:koff} (to account for the planar nature of the membrane), several recent studies \cite{Govern2010, Aleksic2010} have shown that potency data best support a model in which T cells respond to the effective bound time, instead of to the intrinsic bound time or the occupancy alone.

Additionally, a renormalized description of rebinding has quite recently been invoked to explain the observed kinetics of Src homology 2 proteins binding to receptor tyrosine kinases at the membranes of human cells \cite{Oh2012}.  The authors observed a $\sim$$20$-fold decrease in the protein dissociation rate relative to in vitro measurements, which they explained via a reduced effective off-rate of the form of Eqn.\ \ref{eq:koff}, finding that a typical protein locally rebinds to a receptor $\sim$$20$ times.  Furthermore, the authors confirmed that the effective off-rate is dependent on the diffusion constant of the proteins in the cytosol (which explicitly enters Eqn.\ \ref{eq:koff} via $k_D$).
Interestingly, they also observed that the receptors form clusters upon stimulation, further enhancing rebinding, which indeed underscores the intimate link between clustering and rebinding discussed above.

\subsection*{Beyond renormalization: when rebinding causes new behavior}

The success of the renormalized description of rebinding relies strongly on the presence of a timescale separation.  For example, when applying the description to gene regulation in the previous section, it was critical that the rebinding time of the repressor is much faster than the association time of the RNA polymerase.  This timescale separation allows one to treat a group of rapid rebinding events as a single event and to model the downstream dynamics accordingly.  Importantly, however, not all systems prone to rapid rebinding exhibit such a timescale separation.  The spatiotemporal correlations induced by rebinding can then no longer be captured by simple renormalization; instead they introduce a qualitative change in the signaling response.

A well studied system in which rebinding has been shown to cause qualitative signaling changes is the MAPK cascade.  In each layer of the MAPK cascade, a substrate is phosphorylated and dephosphorylated at two sites by a kinase and a phosphatase, respectively.  The network is therefore as shown in the inset of Fig.\ \ref{fig:cluster}C, except that here the kinase ($E_a$), the phosphatase ($E_d$), and the substrate ($S$) are all freely diffusing in three-dimensional space.  In general, two-site modification can be carried out by one of two mechanisms. In a processive mechanism, an enzyme modifies both sites on a substrate molecule before releasing it.  In a distributive mechanism, on the other hand, the enzyme must release the substrate molecule in between modification of the two sites.  Modeling has revealed that the choice of mechanism has important functional consequences.  A distributive mechanism generates an ultrasensitive (very sharp) response because the concentration of doubly phosphorylated substrate depends quadratically on the kinase concentration \cite{Huang1996, Ferrell1996}.  Moreover, a distributive mechanism can lead to bistable behavior of the substrate phosphorylation level if the enzymes are present in limiting amounts \cite{Markevich2004}.

Importantly, these theoretical predictions for the behavior of a distributive network are based on mean-field arguments \cite{Huang1996, Ferrell1996, Markevich2004}, in which the system is assumed to be well mixed.  To test these arguments, particle-based simulations with GFRD were performed on a MAPK pathway \cite{Takahashi2010}, in which it has been shown experimentally that kinases \cite{Burack1997, Ferrell1997} and phosphatases \cite{Zhao2001} can act in a distributive manner. Strikingly, the simulations revealed that both ultrasensitivity and bistability can be lost in regimes in which well-mixed models predict they should be present.  The reason for this loss is rapid rebinding: after a substrate molecule is phosphorylated at its first site and released, it is still close to the kinase and can with high probability rebind to become doubly phosphorylated before diffusing away (Fig.\ \ref{fig:processes}C, right).  In essence, rapid rebinding can turn a distributive mechanism into a processive one, removing the qualitative behaviors associated with distributive modification.  The distinction between a distributive and a processive mechanism then becomes not whether a substrate molecule remains physically connected to its activating enzyme, but rather whether both sites are modified by the same activating enzyme molecule.

Once the effect of rebinding is understood, it becomes clear that the key parameters governing the loss of ultrasensitivity and bistability are the diffusion coefficient and the timescale of enzyme reactivation after the substrate is released (which, in the case of a kinase, could reflect the time for ADP to be exchanged for ATP \cite{Takahashi2010}).  Specifically, the diffusion coefficient must be sufficiently small that rebinding is favored over diffusion into the bulk, and the reactivation time must be sufficiently fast that the enzyme is active once rebinding is attempted.  Indeed, given a sufficiently fast reactivation time, the input-output response of the dual phosphorylation cycle shifts from an ultrasensitive response to a significantly more shallow response as the diffusion coefficient is lowered, ultimately approaching the response of a processive mechanism (Fig.\ \ref{fig:renorm}C).

Furthermore, in the bistable regime, measuring the final fraction of doubly phosphorylated substrate over a range of initial conditions reveals that the system shifts from bistable to monostable as the enzyme reactivation time is reduced (Fig.\ \ref{fig:renorm}D).  This loss of bistability can be understood at a mechanistic level.  When the reactivation time is high, rapid rebinding is suppressed, and the system behaves like it is well mixed.  In a well-mixed system, bistability arises due to enzyme sequestration: if the fraction of inactive substrate is high, the majority of activating enzyme molecules will be occupied by substrate molecules, such that when a substrate molecule gets released in its singly active state, it is more likely to encounter a free deactivating enzyme molecule than a free activating enzyme molecule, thereby maintaining the high fraction of inactive substrate.  A symmetric argument holds for the case when the fraction of doubly active substrate is high, thus supporting bistable behavior.  However, when the reactivation time is low, rapid rebinding is possible.  Then, when a substrate molecule gets released in its singly active state, it can rapidly rebind to the activating enzyme molecule that just released it, even though there are many more free deactivating enzyme molecules in the bulk.  Rapid rebinding thus circumvents the mechanism that supports bistability, resulting in a monostable system.

Both functional changes---the loss of ultrasensitivity and the loss of bistability---are direct results of rapid rebinding, and as such they are not captured by a well-mixed model.  Moreover, they constitute qualitative changes and are therefore not describable by simple renormalization.  Indeed, both effects arise strictly due to molecular spatiotemporal correlations in an otherwise homogeneous system.  When coupled to the MAPK signaling network, these microscopic correlations induce macroscopic changes for the entire cell.

\subsection*{When rebinding can be integrated out and when it cannot}

\begin{figure}
\centering
\includegraphics[width=4in]{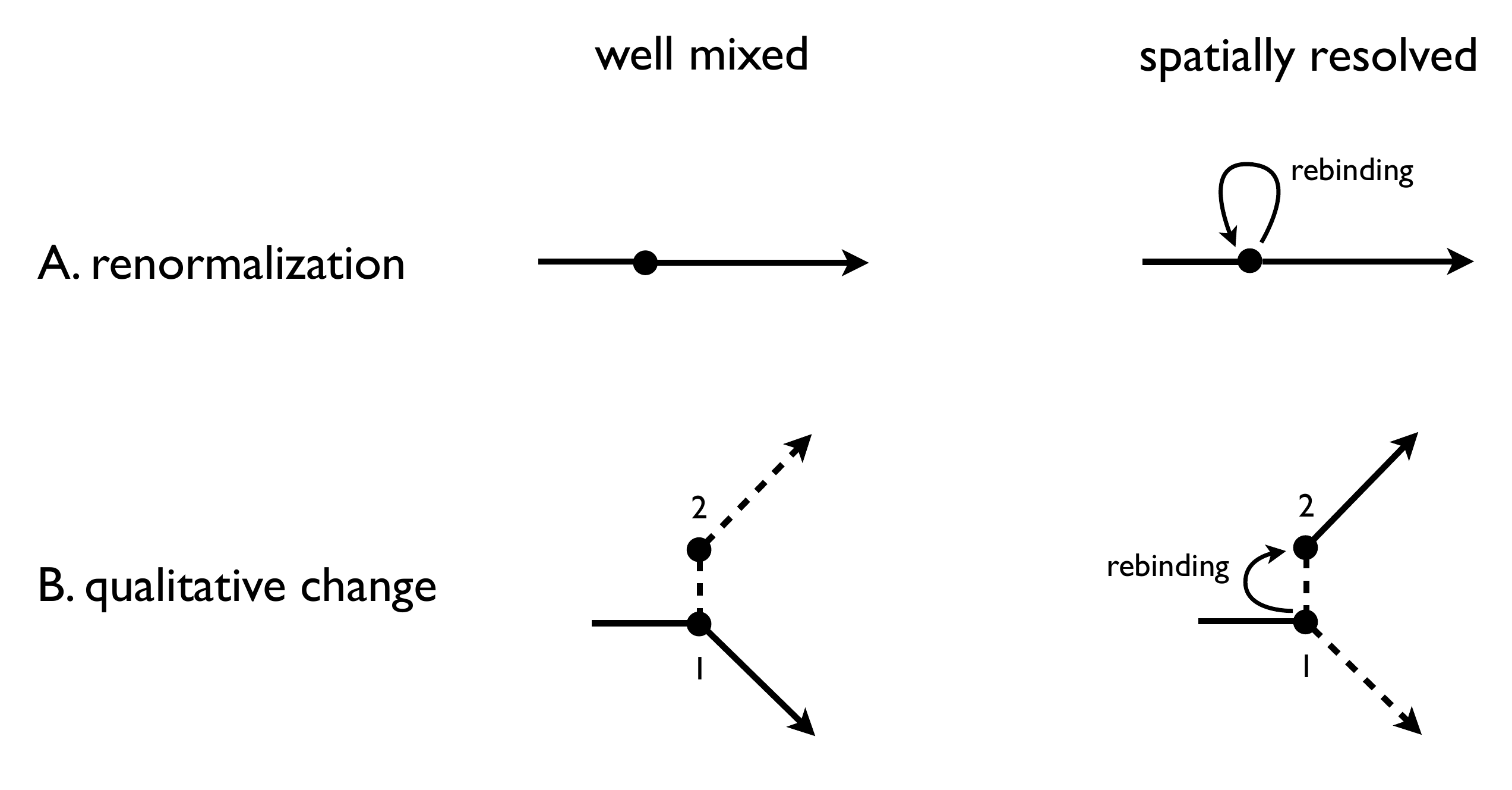}
\caption{Abstract representation of the effects of rebinding.  Dots represent the states of a system, solid arrows represent paths taken by the system in state space (with curved arrows representing rebinding), and dashed arrows represent untaken paths.  (A) Renormalization is possible when rebinding returns the system to the same state and when it does so on a timescale that is faster than that governing the system behavior.  The system then behaves as if it were well mixed, but with renormalized rates.  (B) On the other hand, a qualitative change is induced when rebinding takes the system from one state (1) to another state (2).  Whereas in a well-mixed model the transition from state 1 to state 2 occurs with low probability, in a spatially-resolved model it may occur with high probability due to rebinding.  For example, in the MAPK system, after a kinase has just released its substrate molecule in the singly phosphorylated state (state 1), the substrate molecule either can bind a kinase or a phosphatase from the bulk, leading to ultrasensitivity or bistability (downward arrow), or it can rebind to the kinase that just released it (state 2), leading to a loss of ultrasensitivity or bistability (upward arrow).}
\label{fig:states}
\end{figure}

We are now in a position to understand more generally the distinction between processes in which rebinding can be captured by renormalization and processes in which rebinding induces a qualitative change (Fig.\ \ref{fig:states}).  Renormalization is possible when (i) the state of the system is the same before and after rebinding (Fig.\ \ref{fig:states}A) and (ii) the timescale of rebinding is faster than the timescale that  governs the system dynamics.  Both features are required in order to treat a group of rebinding events as a single effective event, which is the assumption that underlies renormalization.  As we have seen, this type of process is exemplified by gene regulation: the transcription factor--DNA complex is identical before and after a rebinding event, and the rebinding time is much faster than the RNA polymerase association time, making it possible to integrate out the rebinding as if the RNA polymerase were not present.  On the other hand, a qualitative change is induced when rebinding takes the system from one state to another state, in turn accessing a different dynamical pathway (Fig.\ \ref{fig:states}B).  Whereas in a well-mixed system this pathway is suppressed, when rapid rebinding is possible this pathway is allowed, ultimately leading to behavior that is different from the well-mixed prediction.  This type of process is exemplified by distributive phosphorylation: after a kinase has just released its substrate molecule in the singly phosphorylated state, the substrate molecule either can bind a kinase or a phosphatase from the bulk, leading to ultrasensitivity or bistability, or it can rebind to the kinase that just released it, leading to a loss of ultrasensitivity or bistability.

\subsection*{Macromolecular crowding and anomalous diffusion}

The requirement that molecules diffuse toward each other in order to react is compounded in the intracellular environment by the high density of other macromolecules such as DNA and proteins that are not involved in the reaction, a phenomenon termed molecular crowding \cite{Zimmerman1993}.  Molecular crowding is quite significant in cells: the concentration of macromolecules in a prokaryotic cell is typically $300$$-$$400$ g/l \cite{Zimmerman1991}, implying that macromolecules occupy up to $20$$-$$30\%$ of the cellular volume \cite{Ellis2001}.  In fact, in eukaryotic cells, the nucleus is particularly crowded, containing up to $400$ g/l of chromatin alone \cite{Bancaud2009}.  Because crowding can have significant effects on the diffusion and effective reaction rates of molecules, crowding can affect cell signaling in functionally important ways.

Specifically, crowding has two principal effects.  First, it enhances the probability for reactants to be in the bound state, thereby shifting the equilibrium constant of the reaction.  The enhancement is due to an effective attraction between molecules which is often called a depletion interaction or an excluded volume interaction.  The attraction arises directly from the fact that molecules typically take up less total volume in the bound state than in the unbound state, which frees up volume for the surrounding crowder molecules, thus decreasing the free energy of the system by increasing the total entropy \cite{Asakura1954}.  Second, crowding changes the dynamics.  At long length and timescales (e.g.\ length scales larger than that of the spatial heterogeneity in the system), the diffusion of molecules appears normal (i.e.\ the mean squared displacement $\Delta^2$ scales linearly with time $t$), but with an effective diffusion coefficient that is reduced by up to several orders of magnitude \cite{Muramatsu1988}.  On short length and timescales, on the other hand, crowding can lead to anomalous diffusion, or, more precisely, subdiffusion ($\Delta^2 \propto t^\alpha$ with $\alpha < 1$) \cite{Weiss2004, Banks2005, Golding2006, Benichou2010}.  Mechanisms by which crowders induce such subdiffusive motion include forcing a molecule to explore its environment densely (i.e.\ with significant oversampling), such that initial positions take much longer to be forgotten, a type of motion that has been termed compact diffusion \cite{Benichou2010}.

The anomalous diffusion induced by crowding leads to enhanced rebinding \cite{Lomholt2007}.  Indeed, recent analytic work has shown that subdiffusion significantly extends the timescales over which rebinding occurs \cite{Benichou2010, Lomholt2007}, which leads to nontrivial non-exponential waiting time statistics.  However, the transition between anomalous and normal diffusion occurs on a relatively short timescale.  The transition has been measured in vivo, for example with quantum dots in the nuclei of eukaryotic cells, to occur on a timescale of approximately $20$ ms \cite{Bancaud2009}.   Given that this timescale is typically faster than the timescales governing gene expression, and further that the rebinding of transcription factors does not take the system to a new state (Fig.\ \ref{fig:states}A), it has been argued that the effects of crowding on gene expression can be captured by renormalization of the rate constants, with exponential waiting time distributions for association and dissociation \cite{Morelli2011}.  This approach was used to study the effects of crowding on gene networks.  Specifically, Monte Carlo simulations following this argument revealed that the effects of crowding on gene expression are mainly caused by the shift of equilibrium constants rather than the slowing of diffusion \cite{Morelli2011}.  On the other hand, in signaling networks such as the MAPK network, in which rebinding takes the system to a new state (Fig.\ \ref{fig:states}B), the effects of crowding can be nontrivial.  Indeed, crowding-induced rebinding was predicted to contribute to the mechanism by which the distributive double phosphorylation mechanism present in the MAPK network can exhibit a processive response (Fig.\ \ref{fig:renorm}C) \cite{Takahashi2010}.  In fact, subsequent experiments on the MAPK pathway in mammalian cells have confirmed that crowding can indeed convert a distributive mechanism into an effectively processive one \cite{Aoki2011}.

\section*{Outlook}

We have highlighted recent examples of investigations that uncover the mechanisms by which molecular heterogeneity can affect cellular responses.  We have seen that spatiotemporal correlations introduced by molecular-scale structures, such as membrane domains, clusters, scaffolds, and macromolecules, can be amplified by biochemical networks, leading to a changes for the entire cell.  Moreover, we have emphasized that intrinsic spatiotemporal correlations remain present even in macroscopically homogeneous systems, due to the small number of reacting molecules, the finite speed of diffusion, and the crowded cellular environment, and that these intrinsic correlations can be just as potent at inducing cellular changes.  

Several themes emerge from our discussion.  We have seen, for example, in discussing the effects of serial ligation, that while changing the spatial structure (e.g.\ by clustering) can affect the statistics of an equilibrium binding process, coupling to a non-equilibrium network is required to change the mean response.  Further, we have seen that renormalization relies not only on a separation between the rapid timescale of  rebinding and the slower timescale of network dynamics, but also on the property that rebinding returns the system to the same state from which it left.  When rebinding takes the system to a new state, renormalization is not possible; instead rebinding can lead to a qualitatively different response (Fig.\ \ref{fig:states}).  It will be interesting to see how these themes inform future investigations and whether they can be sharpened into more quantitative or more general principles.

Although we have focused on specific examples, the effects of molecular heterogeneity are broad in scope.
The spatial organization of receptors and associated membrane-bound proteins is a key determinant in the detection of external signals by cells as diverse as bacteria \cite{Bray1998} and immune cells \cite{Li2004}.  Furthermore, multisite phosphorylation of the type that shapes the response of the MAPK pathway is ubiquitous in signaling networks, including those that govern bacterial circadian rhythms \cite{vanZon2007}, the timing of the cell cycle in yeast \cite{Nash2001}, and long-term memory in the nervous system \cite{Miller2005}.

Further investigation of molecular heterogeneity and its functional consequences will continue to rely on advances in both experimental and theoretical techniques.  Recent years have seen exciting advances in super-resolution microscopy, including the coordinated application of an excitatory laser pulse with a concentric de-excitatory pulse \cite{Hell1994} and the stochastic activation of single fluorescent molecules \cite{Rust2006, Betzig2006}.  In parallel, advances in particle-based modeling \cite{Andrews2004, Plimpton2005, Takahashi2010} and computational power have allowed progressively more detailed quantitative predictions to be made.  It will be exciting to see what continued advances on both the experimental and theoretical fronts---and the fruitful exchanges between them---will bring for the understanding of how cellular responses are shaped at the molecular level.

\section*{Acknowledgments}
This work is part of the research programme of the Foundation for Fundamental Research on Matter (FOM), which is part of the Netherlands Organisation for Scientific Research (NWO).  We thank Chris Govern for a critical reading of the manuscript.

\bibliographystyle{unsrt}
\bibliography{mugler_tenwolde_references}

\end{document}